\documentclass[12pt]{article}
\usepackage{amsmath,bm,fleqn,amssymb,graphicx}

\newcounter{defcounter}
\usepackage{amsthm}

\begin{document}

\title{Cyclotron Resonance Gain for FIR and THz Radiation in Graphene}
\author{Nightvid Cole and Thomas M. Antonsen Jr.*\\
\multicolumn{1}{p{.7\textwidth}}{\centering\emph{Institute for Research in Electronics and Applied Physics and Department of Physics, University of Maryland, College Park MD 20742}}\\ *also Department of Electrical and Computer Engineering}
\date{}

\maketitle

\subsection*{Abstract}

A cyclotron resonance maser source using low-effective-mass conduction electrons in graphene, if successful, would allow for generation of Far Infrared (FIR) and Terahertz (THz) radiation without requiring magnetic fields running into the tens of Tesla. In order to investigate this possibility, we consider a situation in which electrons are effectively injected via pumping from the valence band to the conduction band using an infrared (IR) laser source, subsequently gyrate in a magnetic field applied perpendicular to the plane of the graphene, and give rise to gain for a FIR/THz wave crossing the plane of the graphene. The treatment is classical, and includes on equal footing the electrons’ interation with the radiation field and the decay in electron energy due to collisional processes. A set of integral expressions is derived by assuming that the non-radiative energy loss processes of the electrons can be adequately represented by a damping force proportional and antiparallel to their momentum. Gain is found even though there is no inversion of the energy distribution function. Gain can occur for electron damping times as short as hundreds of femtoseconds.

\subsection*{I. Introduction}
The conventional gyrotron \cite{paper1}, also known as the cyclotron resonance maser, is a microwave source that makes use of stimulated cyclotron radiation. Stimulated radiation is possible due to the relativistic energy dependence of the gyration frequency of an electron in a uniform magnetic field. According to classical gyrotron theory, a gyrotron will produce radiation at the fundamental or a harmonic of the base
angular velocity of gyration. For a particle of charge $q$ and mass $m$ at a relativistic
factor $\gamma = \left( 1 - v^2/c^2 \right) ^{-1/2} $ and in a uniform magnetic field of strength $B$, this is given by

\begin{align}
\omega = \frac{|qB|}{\gamma m}.
\end{align}

\noindent{For} an electron with relativistic factor $\gamma =1.1$, and $|B|=2 \rm{T}$, this yields an output fundamental (cyclic) frequency of
\begin{align}
f =  \frac {\omega}{2 \pi}\ =\ 50.8\ \rm{GHz}.
\end{align}

It is desired to raise this into the terahertz (THz) range without requiring extraordinarily high magnetic fields of several tens of tesla, which are only available in expensive, large-scale, superconducting, and pulsed, electromagnets\cite{paper2}. This can be accomplished if the effective mass of the electron can be lowered by working with conduction band electrons of a solid material \cite{paper3}, while essentially retaining the principles of gyrotron physics. Consideration of a graphene-based gyrotron is the subject of this paper. By contrast, in many traditional semiconductors, the band structure is more complex, and so is the emission spectrum from transitions between Landau levels. Band structure and emission spectra can be kept simple by using Landau levels of a single species of charge carrier, and graphene can keep effective mass low enough to allow for magnetic fields available at $ T \geq 77 \ \rm{K} $  to suffice. Light-to-heavy-hole lasing, one of the most plausible semiconductor alternatives for producing tunable far infrared oscillators, would require magnetic fields which might pose a problem for operation at $ T \geq 77 \ \rm{K} $, and Germanium Semiconductor-based cyclotron resonance maser (SCRM) sources, although otherwise promising,  require even lower temperatures since only the lowest few Landau levels are involved \cite{paper4} .

Previous work has focused on issues such as population inversion of Landau states in graphene \cite{paper5}, \cite{paper6}, on graphene Landau lasing in the quantum regime \cite{paper7}, \cite{paper8}, \cite{paper9}, on THz gain in optically pumped graphene with no magnetic field \cite{paper10}, THz gain in graphene using dielectric substrates and photonic boundary conditions \cite{paper11}, and femtosecond-scale transient population inversion in optically pumped graphene due to carrier cooling and Auger recombination \cite{paper12}. This paper presents an analysis of the possibility of achieving gain for THz fields in graphene from a semiclassical perspective, and finds that it may be viable if pumped by an appropriate source of electron-hole pairs, such as a mid-infrared laser of suitable strength.  We find gain even though there is no population inversion in the model.  Gain is possible due to a correlation between an electrons energy and its time of “birth”.
Previous work at low, zero, and negative Landau states such as in \cite{paper7} concluded that Auger scattering prevents population inversion from occurring, and thus restricts gain.  However the effect of Auger scattering at large quantum numbers is to contribute to an effective (classical) damping force.  In the classical model, the Coulomb interaction between electrons gives rise chiefly to small-momentum-transfer scattering events and can be thus approximated by a damping force on the electrons as they scatter by small angles and energy shifts from many other electrons successively.   In the absence of a coherent radiation field, this slowing down also leads to a distribution function which decreases monotonically with increasing  energy. However, if the slowing down and interaction with the radiation field are treated consistently we find gain is possible.  While the full features of Auger scattering at low Landau levels require a quantum treatment, the classical treatment should suffice at large quantum numbers when many states are available for electrons to scatter into.

While it may seem paradoxical that gain can occur without population inversion, one should bear in mind that the usual argument linking gain to population inversion assumes a \textit{statistical} mixture (i.e. a completely \textit{incoherent} superposition) of different energy levels. Indeed, the equivalent assumption in the classical picture also results in a conclusion of no gain.  Despite this, the phase bunching that occurs in the gyrating electrons can still give rise to gain when the electron birth times are correlated with their energies, as we will later show. This correlation leads to an energy-dependent gyration phase distribution relative to the phase of the THz field. Returning to the quantum mechanical picture, the assumption of statistically independent energy states does not apply to the case in our model, because a classical gyration phase, reinterpreted in quantum terms of a localized wave packet, is related to the relative phase between neighboring levels in a \textit{coherent} superposition of energy eigenstates (Landau levels) with a common center of gyration. A bunching of classical gyration phases corresponds to the coherence within a quantum superposition of levels. Thus the statistical/incoherent assumption usually invoked in atomic and molecular systems does not hold. A coherence between states has been shown to give rise to gain without inversion in systems with as few as three participating energy levels \cite{paper13}. Thus the presence of a non-inverted population, is not by itself a sufficient condition to show that gain cannot occur.

%\subsection*{Theory}

\begin{figure}[h!]
\centering
\includegraphics[width = 6 in , height = 5.5 in]{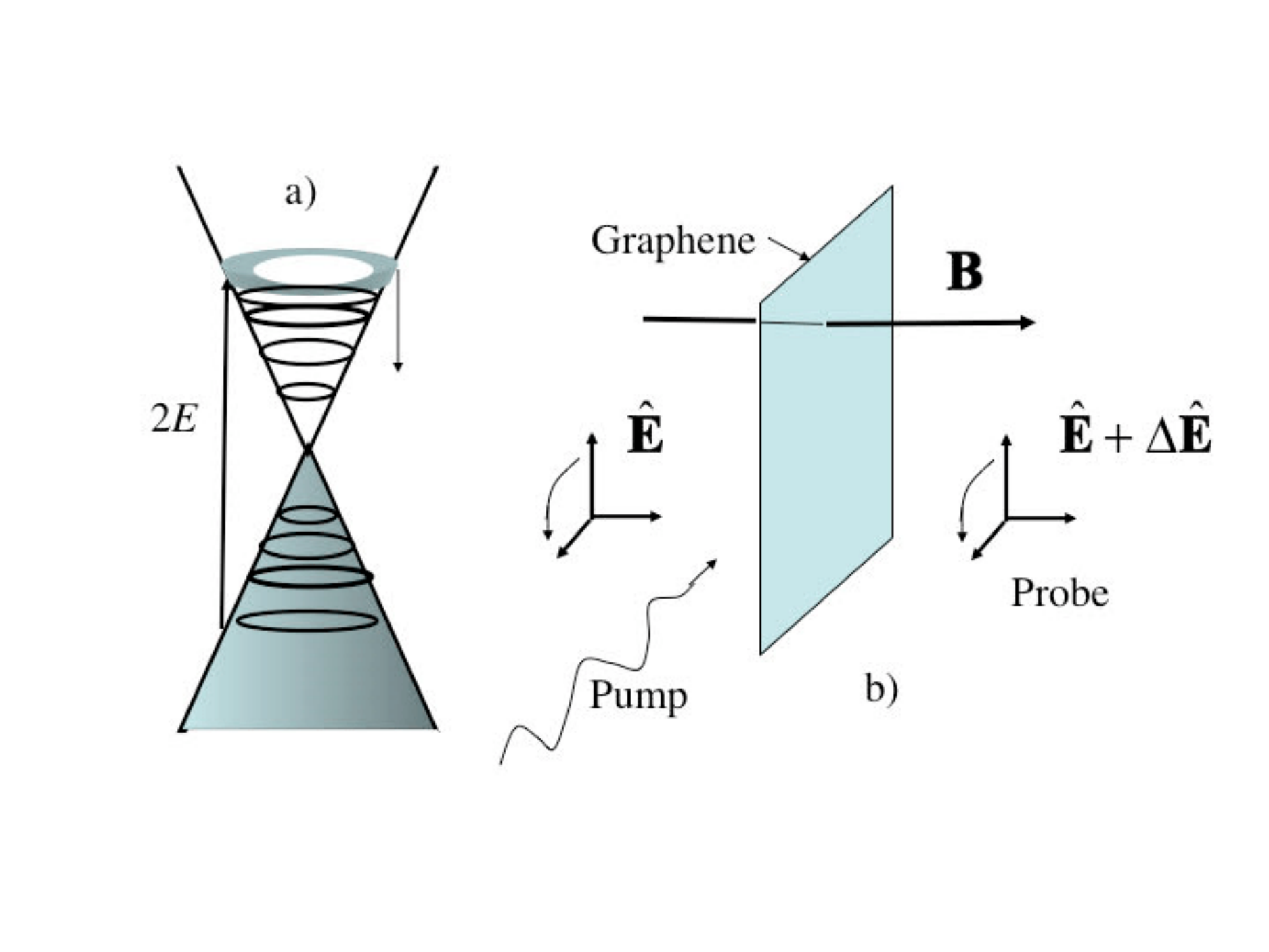}
\caption{a) Schematic of graphene band structure showing valence electrons and electrons excited by a laser pulse with photon energy 2$E$. b) Schematic of configuration analyzed showing orientation of graphene sheet, applied magnetic field, and incident and transmitted probe wave.}
\label{fig:pumping1}
\end{figure}

Undoped graphene has the Dirac band structure \cite{paper3}, equivalent to ultrarelativistic (or massless) electrons, with a band velocity of $ 10^6 \ \rm{m/s}$. The conduction band and the valence band are touching at the Dirac point, as illustrated in Fig. \ref{fig:pumping1}a. The energy-momentum dispersion relation for the massless (zero band gap) band structure is

\begin{align}
E = pc^{\prime},
\end{align}
where $c^{\prime}$ is the band velocity. In the present study, we allow for the possibility of a band gap that could be achieved by doping and doubling the graphene to bilayer graphene \cite{paper14}, but the bandgap is not necessary and is taken to be small. The energy-momentum dispersion relation for the massive (nonzero band gap) band structure is

\begin{align}
E = \sqrt{(pc^{\prime})^2 + (m^{\prime} c^{\prime ^2})^2},
\end{align}
where $c^{\prime}$ is the high-momentum band velocity and $m^{\prime}$ is the effective mass. The band gap energy is $2 m^{\prime} c^{\prime 2}$. Furthermore, we assume that the Fermi level is tuned to $E = 0$, between the valence band and the conduction band. Thus, the electrons can be made to behave in a manner analogous to relativistic electrons in a conventional gyrotron. We also have the analogous relativistic factor $\gamma^{\prime}$ such that the energy is $E = \gamma^{\prime} m^{\prime}c^{\prime ^2}$ and the gyration frequency satisfies $\omega = eB/\gamma^{\prime}m^{\prime} = eBc^{\prime ^2}/E$.

The situation that we consider is illustrated in Fig. \ref{fig:pumping1}b. A sheet of graphene is oriented perpendicular to a uniform magnetic field. The graphene is illuminated by a pump and a probe. The pump beam excites an electron from the valence band to the conduction band as illustrated in Fig. \ref{fig:pumping1}a. The electron loses energy to collisions passing through multiple states of the ideal Hamiltonian while interacting with the wave electric field. The probe beam passes normally through the graphene and experiences gain or loss due to the response of the graphene electrons.

An exploration based on a classical treatment of electron motion in a strong applied magnetic field will be considered here and is valid when electron excitation energy is sufficiently large, such that the electrons responsible for the gain are in high order Landau levels, with index $N \simeq [E^2 - (m^{\prime} c^ {\prime ^2})^2]/(2 \hbar |qB| c^{\prime^2}) >> 1$, where $\hbar$ is the reduced Planck constant. If we work in the low-band-gap limit where $E >> m^{\prime} c^ {\prime ^2}$, this may be written as

\begin{align}
N \simeq 7.6 \times 10^2 \frac{(E[\rm{eV}])^2}{(B[\rm{T}])}.
\label{eqn:new48}
\end{align}

The gyration frequency of an electron with injected energy $E_i$ is given by $\omega_i = eBc^{\prime ^2} / E_i$ with $c^{\prime ^2} = 10^{12} \rm{m^2/s^2}$. Our analysis assumes that the electrons slow down due to collisions and thus, their gyrofrequency changes with time. We will find that maximum gain of the probe occurs due to electrons that have slowed such that their gyration frequency is a factor of about 1.7 times their initial frequency. The maximum gain occurs for frequencies

\begin{align}
f[\rm{THz}] = 0.27 \frac{B[\rm{T}]}{E[\rm{eV}]}.
\label{eqn:new49}
\end{align}

\noindent From Eqs. (\ref{eqn:new48}) and (\ref{eqn:new49}), the level number, frequency and energy relate via
\begin{align}
N = 2.1 \times 10^2 \frac{E[\rm{eV}]}{f[\rm{THz}]}.
\end{align}

\noindent Thus for example, using an 8 T magnetic field as an upper limit, a 500 meV electron excited to the $N$ = $10^{th}$ state will give rise to gain at a frequency of 4 THz. The condition on $N$ is more stringent than simply $N>1$.  This is because, as we will find, gain occurs in narrow bands of frequency $\delta f/f$ of order 0.2.  It is necessary that there be a sufficient number of transitions in this frequency range so that the classical picture involving a superposition of states can apply.  Thus, we can expect to have to consider cases where $N$ is of order 10 or higher.

To treat the situation we consider quantum mechanically one would need to include the effect of many transitions between states.  Recall that the electron energy drops from its initial value to one that is roughly 50 \% lower during the process of slowing down.  Further, as we will find from the classical picture, the electron-field interaction occurs for a finite time, several wave periods.  The quantum wave function describing this would thus involve a coherent superposition of multiple states. The classical approach accounts for these two effects in the limit of a large number of participating states, and provides an answer in terms of simple integrals and figures.

 The organization of this paper is as follows: In section II we derive classical equations for the gain or loss an electromagnetic wave experiences in crossing transversely a layer of graphene in which electrons have been energized. The main result of this analysis is an expression (Eqs.(\ref{eqn:new40}-\ref{eqn:new18})) for the complex gain of the probe wave. This expression is evaluated numerically and the gain is plotted as a function of its independent parameters. The main conclusion is that positive gain can occur if the slowing down time $\tau$ satisfies $\omega_i \tau \textgreater 15$. Section III presents a discussion of issues that are important in realizing gain experimentally and presents sample numbers. Finally, areas for further study are listed.

%\subsection*{Setup}

%See Figure 3.

\subsection*{II. THz gain}

\indent \indent We consider a single atomic layer of graphene that is illuminated by a mid-infrared laser to pump electrons from the valence band to the conduction band, at an initial energy $E_i = \sqrt{(p_i c^{\prime})^2 + (m^{\prime} c^{\prime ^2})^2},$ corresponding to momentum $p_i$ (Figure \ref{fig:pumping1}). A THz wave, to be amplified, is normally incident on the graphene. The electric field of this wave is in the plane of the graphene. The applied static magnetic field is perpendicular to the graphene plane. We take the graphene to be in the $x-y$ plane. As long as the waist diameter of the THz beam is much larger than both the wavelength of the THz radiation and the distances (fast gyration and slow drift, if any) travelled by the electrons during their interaction with the radiation, the system is to a good approximation translationally invariant in both the $x$ and $y$ directions and will be treated as such in this analysis. The electrons are assumed to be governed by classical mechanics once injected into the conduction band by the IR laser.

The THz electromagnetic radiation field can be considered to undergo two processes in passing through the graphene. First, a modification by the current of the excited electrons in the graphene gives gain. Then, a dispersionless, frequency-independent loss due to absorption by the graphene itself. Mirror transmission, and mirror absorption (including large-angle scattering out of the cavity) would occur in a self sustaining oscillator as will be discussed in sec. III.

An electron at time t has a momentum $\textbf{p}(t)$ which may be expressed in Cartesian coordinates as
\begin{align}
\textbf{p}(t) = p_x(t)\hat{x} + p_y(t)\hat{y}.
\end{align}
Introducing polar coordinates $p(t)$ and $\theta(t)$, this becomes
\begin{align}
\textbf{p}(t) = p(t)[\cos \theta (t)\hat{x} + \sin \theta (t)\hat{y}].
\end{align}
Differentiating the latter with respect to time, and using the Lorentz force relation $\dot{\textbf{p}} = q\big[ \textbf{E} + \textbf{v} \times \textbf{B} \big]$, we obtain the pair of equations

\begin{align}
\dot{p} = -e[E_x \cos \theta + E_y \sin \theta],
\end{align}
and

\begin{align}
p \dot{\theta} = -e[-E_x \sin \theta + E_y \cos \theta] + \frac{p \omega_L}{\gamma ^{\prime}},
\end{align}
\
where
\
\begin{align}
\omega_L \equiv \frac{eB}{m^{\prime}},
\label{eqn:new53}
\end{align}
and
\begin{align}
\gamma^{\prime} \equiv \sqrt{1 + \left( \frac{p}{m^{\prime}c^{\prime}} \right)^2}.
\end{align}
\
The above, however, apply only to an idealized case with no scattering of electrons by phonons or inhomogeneities/defects in the graphene. Both elastic and inelastic scattering can occur. A simple model which treats inelastic scattering events of primarily small, longitudinal momentum transfers as an overall damping on the electron momentum causing it to decay exponentially with time constant $\tau$ will be used. The slowing down model produces Drude-like dissipation with a real part of mobility, in the magnetic field free case, that scales as $[1+(\omega \tau)^2]^{-1}$. (A modification of this analysis which incorporates an electron removal process such as large-angle scattering would simply include an extra factor of $e^{-(t - t_B)/ \bar{\tau}}$, where $\bar{\tau}$ is a time constant of removal, inside the integral in Eq. (\ref{eqn:new5}) which would carry directly into the outermost integral of Eq. (\ref{eqn:new40}).) Thus, we are excluding pitch angle and energy diffusion processes, with the rationale being that for superthermal electrons, damping should dominate. In this case, the equation for $\dot{p}$ becomes

\begin{align}
\dot{p} = -e[E_x \cos \theta + E_y \sin \theta] - \frac{p}{\tau},
\label{eqn:new1}
\end{align}

\noindent{while} that for $\dot{\theta}$ is now

\begin{align}
\dot{\theta} = -p^{-1}e[-E_x \sin \theta + E_y \cos \theta] + \frac{\omega_L}{\gamma^{\prime}}.
\label{eqn:new2}
\end{align}

\noindent{Now} we transform to a rotating frame with angular velocity $\omega$ (which will be the frequency of the electromagnetic wave) and initial angle $\phi_0$ by introducing the variable $\overline{\theta} = \theta - \omega t - \phi_0$. With this transformation Equation (\ref{eqn:new1}) becomes

\begin{align}
\dot{p} = -e \left[ \hat{E} e^{i \overline{\theta}} + c.c. \right] - \frac{p}{\tau},
\label{eqn:new6}
\end{align}
and Eq. (\ref{eqn:new2}) becomes

\begin{align}
\dot{\overline{\theta}} = \frac{\omega_L}{\gamma^{\prime}} - \omega - ep^{-1} \left[ i \hat{E} e^{i \overline{\theta}} + c.c. \right],
\label{eqn:new7}
\end{align}

\noindent where

\begin{align}
\hat{E} = \frac{\left( E_x - i E_y \right)}{2} e^{i \omega t}.
\label{eqn:new4}
\end{align}

We note that in the absence of a coherent THz field the momentum relaxation term in Eq. (16) leads to a distribution function that scales with momentum as $\tau / p^2$, and is thus not inverted. We will solve this equation system subject to the following initial conditions. Electrons are injected into the conduction band with energy $E_i = \gamma_i^{\prime} m^{\prime} c^{\prime ^2}$ with $\gamma_i^{\prime} = \left( 1+p_i^2/\left( m^{\prime}c^{\prime} \right) ^2 \right) ^{1/2}$ and with initial momentum angle $\overline{\theta}$ uniformly distributed in the interval $[0,2 \pi]$. Further, each electron has a birth time $t_B$ at which $p = p_i$ and $\overline{\theta} = \overline{\theta}_0$, which we will take to be uniformly distributed. Solutions are then parameterized as follows,
\begin{align}
\begin{matrix}
p = p(t;\overline{\theta}_0,t_B) \\ \overline{\theta} = \overline{\theta}_0 + \Delta \overline{\theta}(t;\overline{\theta}_0,t_B),
\end{matrix}
\end{align}
where $p(t_B;\overline{\theta}_0,t_B) = p_i$ , $\Delta \overline{\theta} (t_B;\overline{\theta}_0,t_B) = 0$.

The electric field, being a pure radiation field, satisfies the driven wave equation

\begin{align}
- \nabla^2 \textbf{E} + \frac{1}{c^2} \frac{\partial^2 \textbf{E}}{\partial t^2} = -\mu_0 \frac{\partial \textbf{J}}{\partial t},
\end{align}

\noindent where $\textbf{J}$ is the conduction electron current density, and we assume the radiation waist is sufficiently large so that we may take $\nabla \cdot \textbf{J} \simeq 0$, and hence $\nabla \cdot \textbf{E} \simeq 0$.

\noindent Consider an electromagnetic wave of the form

\begin{align}
\textbf{E} = E_x\left(t - \frac{z}{c},z\right)\hat{x} + E_y\left(t - \frac{z}{c},z\right)\hat{y},
\end{align}

\noindent which is propagating in the $+z$  direction. Making the substitution of variables
\begin{equation*}
\begin{split}
\overline{t} &\equiv t-\frac{z}{c} \\ \partial / \partial t &= \partial / \partial \overline{t} \\ \left( \frac{\partial}{\partial z} \right)_{old} &= \left( \frac{\partial}{\partial z} \right)_{new} - \frac{1}{c} \frac{\partial}{\partial \overline{t}}.
\end{split}
\end{equation*}

\noindent The wave equation becomes

\begin{align}
\left( \frac{2}{c} \frac{\partial}{\partial \overline{t}} \frac{\partial}{\partial z} - \frac{\partial^2}{\partial z^2} \right) \left( E_{x}(\overline{t},z) \hat{x} + E_{y}(\overline{t},z) \hat{y} \right) = -\mu_0 \frac{\partial \textbf{J}}{\partial \overline{t}}.
\end{align}

\noindent Discarding the term $ \partial^2 / \partial z^2 $ (negligible graphene reflection) and integrating with respect to  $\overline{t}$ gives

\begin{align}
\frac{2}{c} \frac{\partial}{\partial z} \left( E_{x}(\overline{t},z) \hat{x} + E_{y}(\overline{t},z) \hat{y}  \right) = - \mu_0 \textbf{J},
\end{align}

\noindent where we have set the constant of integration that corresponds to a zero-frequency component of radiation to zero. This result can in turn be integrated with respect to z, from just before radiation passes through graphene at $z = 0$ to just after. The result is

\begin{align}
\frac{2}{c} \Delta (\textbf{E}) = -\mu_0 \int_{z \rightarrow 0^-}^{z \rightarrow 0^+} \textbf{J} \left( z \right) dz,
\label{eqn:new3}
\end{align}

\noindent where

\begin{align}
\Delta (\textbf{E}) \equiv (E_{x}(\overline{t},z) \hat{x} + E_{y}(\overline{t},z) \hat{y})\big{|}_{z \rightarrow 0^+} - (E_{x}(\overline{t},z) \hat{x} + E_{y}(\overline{t},z) \hat{y})\big{|}_{z \rightarrow 0^-},
\end{align}

\noindent which represents the change in electric field components on transiting the graphene. We now evaluate the integrated current density on the RHS of Eq. (\ref{eqn:new3}).

\
\

Each conduction electron with velocity $\textbf{v}(t;t_B,\overline{\theta_0}) = \textbf{p}(t;t_B,\overline{\theta_0})/(\gamma(t;t_B,\overline{\theta_0})m^{\prime})$ and birth time $t_B < t,$ will contribute to the current density. If the pumping IR laser excites electrons to the conduction band at a rate $\dot{n}$ , where $\dot{n}$ has units $m^{-2} s^{-1}$, the electric field change can be thus cast in terms of the velocity of an electron given its history as

\begin{align}
\Delta \textbf{E} = -\frac{c\mu_0}{2} \int_{z \rightarrow 0^-}^{z \rightarrow 0^+} \textbf{J}(z) dz   = \frac{eZ_0\dot{n}}{2} \int_{-\infty}^t dt_B \left\langle \textbf{v}(t;t_B,\overline{\theta}_0) \right\rangle_{\overline{\theta_0}},
\end{align}

\noindent where $Z_0 = \sqrt{\mu_0/ \epsilon_0}$ is the impedance of free space and we have used $c = 1/\sqrt{\mu_0 \epsilon_0}$.

\noindent This gives the jump in Cartesian components of the electric field the wave experiences due to the conduction electrons. The change in the complex amplitude $\hat{E}$ defined by Eq. (\ref{eqn:new4}) can then be expressed
\
\
\begin{align}
\Delta \hat{E} = \frac{eZ_0\dot{n}}{4} \int_{-\infty}^{t} dt_B \left\langle \left( v_x - i v_y \right) e^{i \omega t} \right\rangle = \frac{eZ_0\dot{n}}{4m^{\prime}} \int_{-\infty}^{t} dt_B \left\langle \frac{pe^{-i\overline{\theta}}}{\gamma} \right\rangle_{\overline{\theta}_0}.
\label{eqn:new5}
\end{align}

We now seek to calculate the conditions under which the growth of a wave with prescribed frequency $\omega$ due to the interaction with the excited graphene electrons can overcome the combined losses due to absorption in the graphene and mirrors and transmission through the mirrors. To this end we assume that the electric field is oscillating sinusoidally, $\textbf{E}(t) = Re\{\tilde{E}e^{-i \omega t}\}$, and is sufficiently small that the equations of motion can be linearized. We now proceed to perform this linearization. We write the electron momentum as the sum of the field free component, with subscript ``0'', and a first order in electric field perturbation with subscript ``1'',

\begin{align}
p \left( t-t_B \right) = p_0 \left( t-t_B \right) + p_1 \left( t-t_B \right),
\end{align}
and
\begin{align}
\Delta \overline{\theta} \left( t-t_B \right) = \Delta \overline{\theta}_0 \left( t-t_B \right) + \Delta \overline{\theta}_1 \left( t-t_B \right).
\end{align}

\noindent The field free solutions satisfy Eqs. (\ref{eqn:new6}) and (\ref{eqn:new7}) with $\hat{E}$ set to zero,

\begin{align}
p_0 \left( t-t_B \right) = p_i e^{- \left(t-t_B \right) / \tau },
\label{eqn:new42}
\end{align}
and
\begin{align}
\Delta \overline{\theta}_0 \left( t-t_B \right) = \int_{t_B}^t dt^{\prime} \left( \frac{\omega_L}{\gamma_0 \left( t^{\prime}\right)} - \omega \right) = \left( \omega_L - \omega \right) \left( t - t_B \right) - \omega_L \tau ln \left( \frac{\gamma_i + 1}{\gamma_0 \left( t \right) + 1} \right),
\label{eqn:new14}
\end{align}

\noindent and

\begin{align}
\gamma_0 \left( t-t_B \right) = \sqrt{1 + \frac{p_0^2 \left( t-t_B \right)}{m^{\prime 2} c^{\prime 2}}},
\label{eqn:new20}
\end{align}
is the relativistic factor of an electron as it slows down, and
\begin{align}
\gamma_i = \sqrt{1 + \frac{p_i^2}{m^{\prime 2} c^{\prime 2}}} = \gamma_0 \left( t = t_B \right),
\label{eqn:new54}
\end{align}
is the initial relativistic factor.

In the first order equations the electric field appears. If we take $\omega$ to be the angular frequency of the (radiation) field, the quantity $\hat{E}$ defined in (\ref{eqn:new4}) will have a steady component and a component oscillating at $2 \omega$. (The latter should be negligible with circular polarization.) In first order these act independently, so we take $\hat{E}$ to be steady. The first order equations for electron motion are written
\begin{align}
\frac{d}{dt} p_1 = -\frac{p_1}{\tau} - \left( e \hat{E}e^{i \overline{\theta}_0 + i \Delta \overline{\theta}_0} + c.c. \right),
\label{eqn:new8}
\end{align}
and
\begin{align}
\frac{d \Delta \overline{\theta}_1}{dt} = -\frac{\omega_L}{\gamma_0^2} \frac{d \gamma_0}{dp_0}p_1 - \frac{e}{p_0} \left( i \hat{E} e^{i \overline{\theta}_0 + i \Delta \overline{\theta}_0} + c.c. \right).
\label{eqn:new9}
\end{align}
We note from Eqs. (\ref{eqn:new8}) and (\ref{eqn:new9}) that the dependence of the momentum variables $p_1$ and $\Delta \overline{\theta}_1$ on the birth phase $\overline{\theta}_0$ can be separated according to $p_1 = \hat{p}_1 (t-t_B) e^{i \overline{\theta}_0} + c.c.$ and $\overline{\theta}_1 = \Delta \hat{\overline{\theta}}_1 (t-t_B) e^{i \overline{\theta}_0} + c.c.$ . The complex amplitudes $\hat{p_1}$ and $\Delta \hat{\overline{\theta}}_1$ then satisfy

\begin{align}
\frac{d \hat{p}_1}{dt} = -\frac{ \hat{p}_1}{\tau} - e \hat{E} e^{i \Delta\overline{\theta}_0(t-t_B)},
\label{eqn:new25}
\end{align}
and
\begin{align}
\frac{d \Delta \hat{\overline{\theta}}_1}{dt} = - \frac{\omega_L}{\gamma_0^2} \frac{d \gamma_0}{dp_0} \hat{p}_1 - i \frac{e}{p_0} \hat{E} e^{i \Delta \overline{\theta}_0(t-t_B)} ,
\label{eqn:new26}
\end{align}
with the initial conditions $\hat{p_1}(t = t_B) = \hat{\overline{\theta}_1}(t = t_B) =0.$

\noindent Equations (\ref{eqn:new25}) and (\ref{eqn:new26})  can be integrated giving

\begin{align}
\hat{p}_1 = -e\hat{E}\tau_A(t-t_B),
\label{eqn:new10}
\end{align}
where

\begin{align}
\tau_A (t) = e^{-t/ \tau} \int_0^t dt^{\prime} e^{ t^{\prime}/ \tau + i \Delta \overline{\theta}_0(t^{\prime})},
\label{eqn:new16}
\end{align}

\noindent and
\\
\begin{align}
\Delta \hat{\overline{\theta}}_1(t-t_B) = e \hat{E} \left[ \int_{t_B}^{t} dt^{\prime} \left( \frac{\omega_L}{\gamma_0^3} \frac{d \gamma_0}{d p_0} \tau_A (t^{\prime}-t_B) \right) - \frac{i}{p_i} \tau_A (t-t_B) \right]
%\hat{\overline{\theta}_1} e^{i \overline{\theta}_0} + c.c.,
\label{eqn:new11}
\end{align}

\noindent Next we linearize Eq. (\ref{eqn:new5}), which to first order gives

\begin{align}
\Delta E = \frac{eZ_0\dot{n}}{4m^{\prime}} \int_{-\infty}^t dt_B \left( \left( \frac{1}{\gamma_0} - \frac{p_0}{\gamma_0^2} \frac{d \gamma_0}{d p_0} \right) \hat{p}_1 - i \frac{p_0}{\gamma_0} \Delta \hat{\overline{\theta}}_1      \right) e^{-i \Delta \overline{\theta}_0}.
\label{eqn:new12}
\end{align}

\noindent Upon substituting Eq. (\ref{eqn:new10}) and (\ref{eqn:new11}) in (\ref{eqn:new12}) and letting $\hat{t} = t - t_B$ such that $dt_B = -d \hat{t}$ we find for the increment in field amplitude

\begin{align}
\Delta E = - \frac{e^2 Z_0\dot{n}}{4m^{\prime}} \hat{E} \left( \int_0^{\infty} d \hat{t} e^{-i \Delta \overline{\theta}_0 (\hat{t}) }  \left( \left( \frac{1}{\gamma_0^3(\hat{t})} + \frac{e^{- \hat{t} / \tau}}{\gamma_0(\hat{t})} \right) \tau_A (\hat{t}) + i \frac{p_0}{\gamma_0} \int_0^{\hat{t}} dt^{\prime} \left( \frac{\omega_L}{\gamma_0^2} \frac{d \gamma_0}{d p_0} \right) \tau_A (t^{\prime}) \right) \right),
\label{eqn:new40}
\end{align}

\noindent or simply

\begin{align}
\frac{\Delta \hat{E}}{\hat{E}} = R \left( G-L \right),
\label{eqn:new13}
\end{align}

\noindent where

\begin{align}
R = \frac{e^2 Z_0\dot{n} \tau^2 c^{\prime ^2}}{4 E_i},
\label{eqn:new50}
\end{align}

\noindent is the dimensionless pumping rate, and

\begin{align}
L = \gamma_i \int_0^{\infty} \left( \frac{1}{\gamma_0^3} + \frac{e^{- \hat{t} / \tau}}{\gamma_0} \right) \frac{\tau_A (\hat{t}) e^{-i \Delta \overline{\theta}_0 (\hat{t})}}{\tau^2} d\hat{t},
\label{eqn:new39}
\end{align}

\noindent is a loss term representing absorption of THz by the energetic electrons,

\begin{align}
G = - \gamma_i \int_0^{\infty} \frac{i p_0}{\tau^2 \gamma_0} \int_0^{\hat{t}} dt^{\prime} \left( \frac{\omega_L}{\gamma_0^2} \frac{d \gamma_0}{d p_0} \right) \tau_A (t^{\prime}) e^{-i \Delta \overline{\theta}_0 (\hat{t})} d\hat{t},
\label{eqn:new18}
\end{align}

\noindent is a potential gain term due to gyrophase bunching that allows the THz fields to be amplified. The real part of Eq. (\ref{eqn:new13}) describes the change in the magnitude of the electric field, while the imaginary part describes the change in phase.

Equation (\ref{eqn:new13}) describes the gain or loss the wave experiences on transmission through the graphene. Spontaneous oscillations can grow only for frequencies for which $g \equiv Re(G-L) > 0$, and physical gain of the entire system requires that the pumping rate $\dot{n}$ must be made large enough to overcome transmission, absorption, and scattering losses at the cavity's mirrors and the intrinsic absorption of energy by the valence band electrons of the graphene itself. The precise conditions leading to system gain ($R g \textgreater \ell,$ where $\ell$ represents all losses) will be addressed in the discussion section. For now, we will focus only on the conditions under which $g \textgreater $ 0. The terms ``gain'' and ``dimensionless gain'' in this section, when not otherwise specified, will refer to $g = Re(G-L)$.

The functions G and L have been defined so that they are dimensionless by normalizing to the slowing down time squared. Since G has an extra time integration, we expect it to be larger than L when $\omega_i \tau  \gg 1$. To further characterize the gain we introduce the following parameters: the initial gyration frequency normalized by the slowing down time,

\begin{align}
\omega_i \tau = \omega_L \tau / \gamma_i = \frac{\tau[\rm{ps}] B[\rm{T}]}{E[\rm{eV}]},
\label{eqn:new55}
\end{align}

\noindent the half bandgap energy normalized to the initial energy

\begin{align}
m^{\prime}c^{\prime ^2}/E_i = \gamma_i^{-1},
\label{eqn:new57}
\end{align}

\noindent and the frequency normalized to the initial gyration frequency,

\begin{align}
\omega / \omega_i = \gamma_i \omega / \omega_L.
\label{eqn:new56}
\end{align}

\begin{figure}[h!]
\centering
\includegraphics[width = \columnwidth , height = 5.5 in]{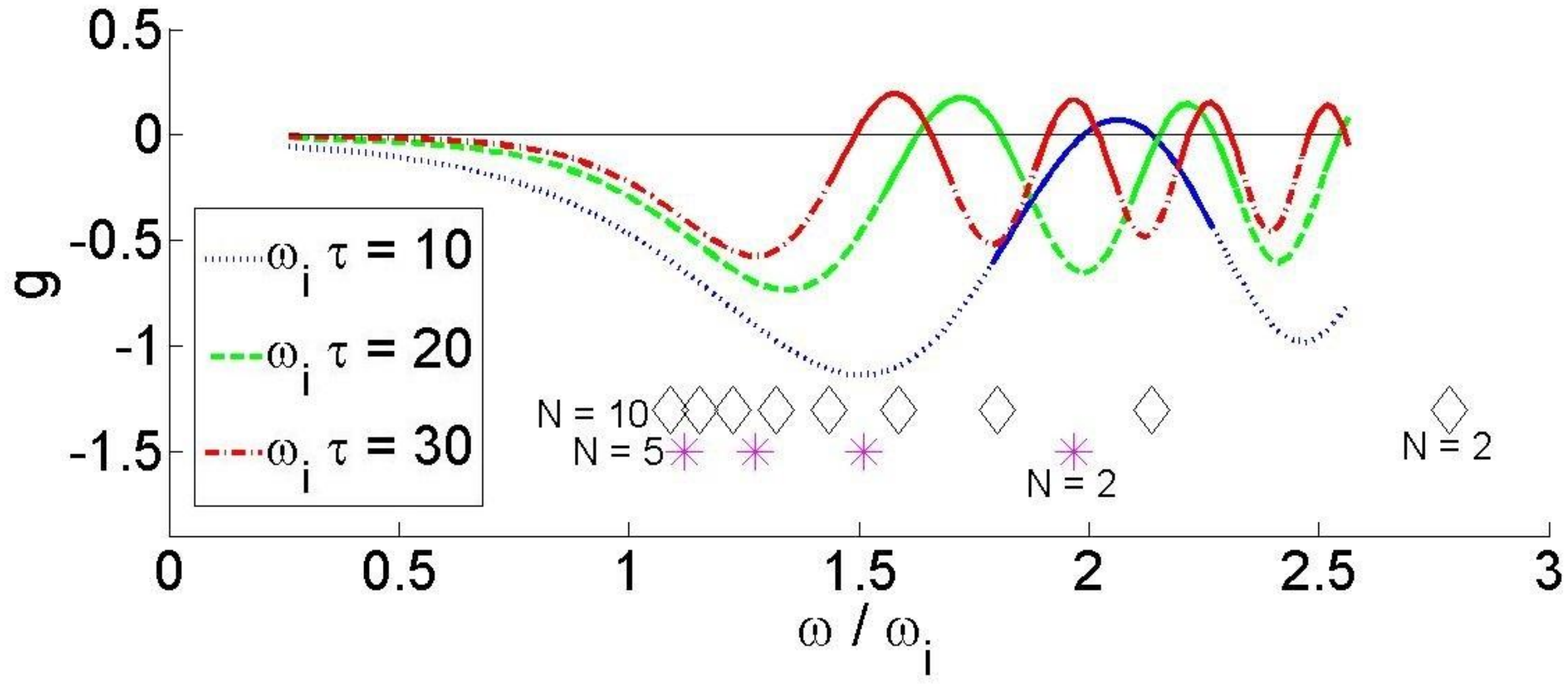}
\caption{Normalized gain versus normalized frequency for several dimensionless slowing down times. For this plot the normalized half bandgap energy is $m^{\prime} c^ {\prime ^2}/E_i = \gamma_i^{-1}$ = 0.00585. The solid portions of each curve indicate where $\cos(\phi_R + \pi / 4) < 0$, where $\phi_R$ is defined in Eqs. (\ref{eqn:new41}) and (\ref{eqn:new44}). Diamonds and stars represent Landau transition energies for 3 T magnetic field, with the N = 10 (diamonds) and N = 5 cases (stars) arising at electron energies of 199 and 140 meV, respectively. g = Re(G-L) where G and L are defined by Eqs. (\ref{eqn:new39}) and (\ref{eqn:new18}), $\tau$ is the time scale of damping of electron motion, and the independent variables are explained by Eqs. (\ref{eqn:new53}), (\ref{eqn:new54}), (\ref{eqn:new55}), and (\ref{eqn:new56}).}
\label{fig:wiggles1}
\end{figure}

Plots of the real part of the dimensionless gain function versus frequency for several slowing down times are shown in Fig. \ref{fig:wiggles1}. The dependence of $g$ on frequency can be characterized as having a slowly varying average part (which is negative) and a superimposed rapidly varying part, which leads to intervals of frequency where gain is positive. Also shown in \ref{fig:wiggles1} as symbols are the frequencies associated with transitions between adjacent Landau levels for two cases: one in which the electrons are initially excited to the fifth Landau level, and one in which they are excited to the tenth.  The classical limit should apply if these symbols are dense enough so that there spacing can resolve the gain curve.  
The origin of the intervals of positive gain is explained as follows. Electrons are injected with initial energy, $E_i$. As an electron slows down, its resonant frequency increases. This means the horizontal axis of Fig. (\ref{fig:wiggles1}) also corresponds to time since birth of the electrons contributing to gain or loss at that frequency. The gain will then show oscillations with frequency corresponding to numbers of integer wave periods since birth. This can be shown as follows. The dependence of gain or loss on frequency enters Eq. (\ref{eqn:new13}) through the phase $\Delta \overline{\theta}_0 (t-t_B)$ defined in Eq. (\ref{eqn:new14}). This phase is a rapidly varying function of time on the scale $t^{\prime} / \tau \sim 1$, except for the interval of time when $\omega_L / \gamma (t-t_B) \simeq \omega$. Expanding the time dependence of the phase around its stationary time, we have (Subscript ``$R$'' refers to resonance)

\begin{align}
\Delta \overline{\theta}_0 \left( t^{\prime} \right) \simeq \phi_{R} + \frac {1}{2} \dot{\Omega} \left( t^{\prime} - t_{R}\right) ^2,
\label{eqn:new52}
\end{align}

\newpage
\

\noindent where $t_{R}$ is defined by $d \Delta \overline{\theta}_0 / dt^{\prime} = 0,$

\begin{align}
\frac{\omega_L}{\gamma_0 \left( t_{R}\right)} - \omega = 0,
\end{align}

\noindent with

\begin{align}
\phi_{R} = \int_{t_B}^{t_{R}} dt^{\prime} \left( \frac{\omega_L}{\gamma_0 \left( t^{\prime}\right)} - \omega \right),
\label{eqn:new41}
\end{align}

\noindent and

\begin{align}
\dot{\Omega} = \frac{d}{dt^{\prime}} \frac{\omega_L}{\gamma_0 \left( t^{\prime} \right)} \Bigr|_{t_{R}} = \frac{p_0^2 \omega_L}{\tau \gamma_0^3 \left( m^{\prime} c^{\prime} \right)^2} \Bigr|_{t_{R}}.
\end{align}

%\newpage \
%\
The smooth part of the gain versus frequency curve comes from the contributions to the integrals in Eq. (\ref{eqn:new18}) from $\tau \simeq t_{R}$ with the additional approximation that the lower limit of the time integrals in Eqs. (\ref{eqn:new16}) and (\ref{eqn:new40}) is taken to be $\tau \rightarrow -\infty$. The rapid oscillations are due to the fact that the endpoint is in fact $\tau = 0$, not $\tau \rightarrow -\infty$. These oscillations thus track the resonant phase $\phi_{R}$ defined in Eq. (\ref{eqn:new41}). The integral in (\ref{eqn:new41}) can be evaluated by switching from $t^{\prime}$ as the integration variable to $\gamma_0 (t-t^{\prime})$ defined through (\ref{eqn:new20}) and (\ref{eqn:new42}). The result is an expression for the resonant phase as a function of frequency,

\begin{align}
\phi_{R} \left( \omega / \omega_L \right) = \frac{\omega_L \tau}{2} \left\{ \ln \left[ \left( \frac{\gamma_i - 1}{\gamma_i + 1} \right) \left( \frac{\omega_L / \omega + 1}{\omega_L / \omega - 1} \right) \right] - \frac{\omega}{\omega_L} \ln \left[ \frac{\gamma_i^2-1}{(\omega_L/ \omega)^2-1} \right] \right\}.
\label{eqn:new44}
\end{align}

The quantity $\phi_{R}$ corresponds to $2 \pi$ times the number of wave periods that elapse between the birth of an electron at $\gamma_i$ and the time it slows down to $\gamma_0 = \omega_L / \omega$. To illustrate its importance we have modified the curves in Fig. (\ref{fig:wiggles1}) such that the curves are solid if $\cos(\phi_{R} + \pi /4) < 0$ and dashed if $\cos(\phi_{R} + \pi /4) > 0$. As can be seen, positive gain occurs only on solid portions of each curve. The origin of the $\pi /4$ phase shift, as well as a more rigorous explanation of the dependence of gain on resonance phase, is presented in the Appendix.

\begin{figure}[h!]
\centering
\includegraphics[width = \columnwidth , height = 5.5 in]{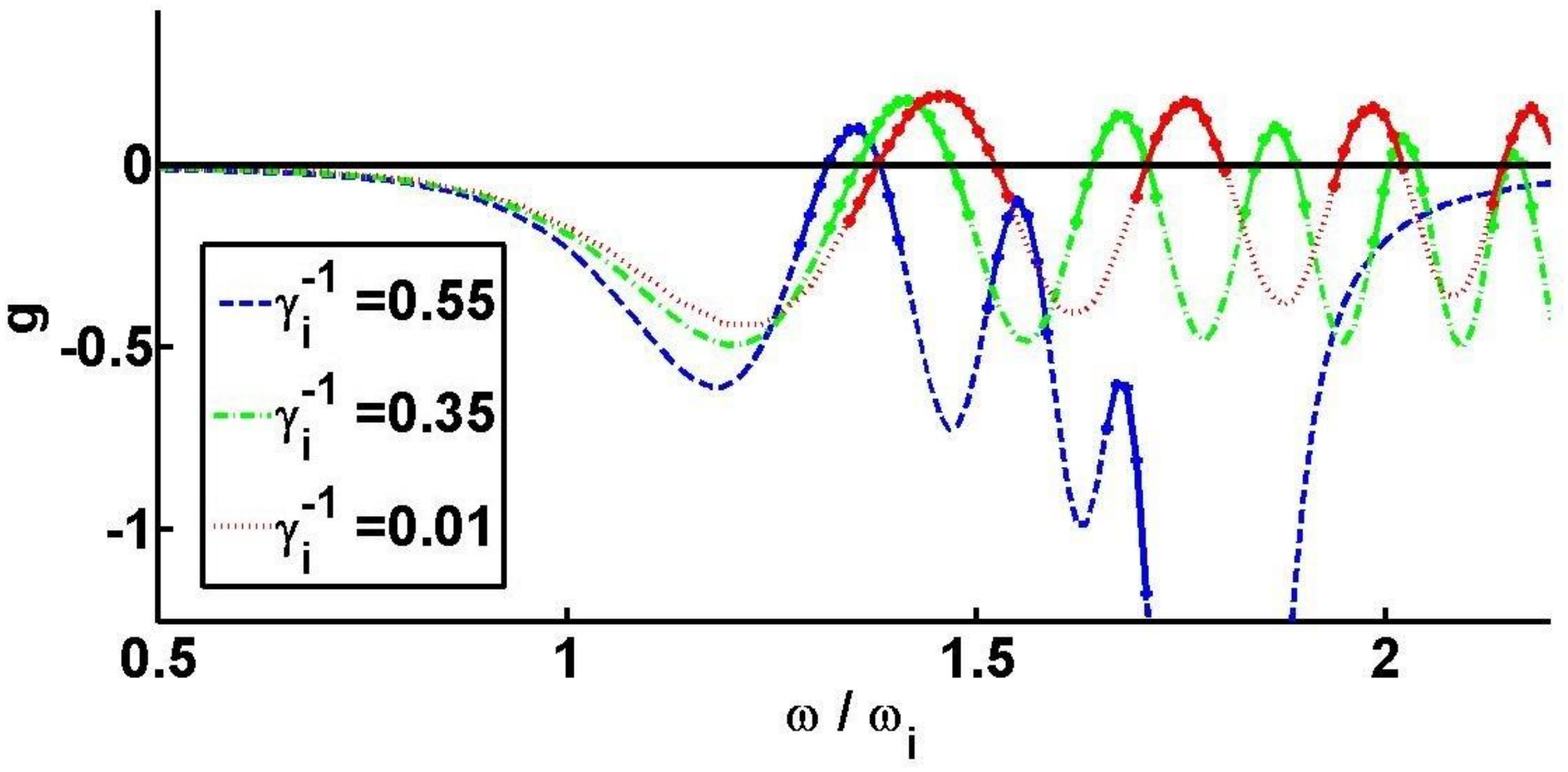}
\caption{Normalized gain versus normalized frequency (same variables as in Fig. \ref{fig:wiggles1}) for several values of normalized half band gap energy. The normalized slowing down time is $\omega_i \tau$ = 44.1 for all curves shown. The curves represent 3 selected half-band-gap energies $\gamma_i^{-1}$ (see Eq. (\ref{eqn:new57})). The solid portions of each curve indicate where $cos(\phi_R + \pi / 4) < 0$, where $\phi_R$ is defined in Eqs. (\ref{eqn:new41}) and (\ref{eqn:new44}).}
\label{fig:bandgapvariation1}
\end{figure}

The effect of varying normalized half bandgap energy is shown in Fig. (\ref{fig:bandgapvariation1}) where normalized gain is plotted vs. normalized frequency for three values of bandgap energy and fixed normalized slowing down time $\omega_i \tau$ = 44.1.

\newpage\

We see from Fig. \ref{fig:bandgapvariation1} that the first gain peak is insensitive to the normalized half bandgap energy once $\gamma_i^{-1}$ is small. Dips in gain occur at frequencies corresponding to the cyclotron resonance at the half bandgap energy $\omega / \omega_i = \gamma_i$. When electrons decrease their energy to the half bandgap value the gyration frequency becomes energy independent, and the negative mass effect responsible for cyclotron resonance gain no longer is possible.

%\begin{figure}[h!]
%\centering
%\includegraphics[width=.995\textwidth]{GminusL_v2.eps}
%\caption{Plots of dimensionless gain $G - L$ as a function of dimensionless frequency $\omega / \omega_L$, for dimensionless initial momentum $ p_i/(m^{\prime}c^{\prime}) = 2.375$ and for three values of the dimensionless decay time constant $\omega_L \tau$. Note that the initial dimensionless gyrofrequency $1/ \gamma_i \simeq 0.388$ in this case. The solid portions of each curve indicate where $cos(\phi_R + \pi / 4) < 0$, where $\phi_R$ is defined in Eqs. (\ref{eqn:new41}) and (\ref{eqn:new44}). NOTE THAT THE Y AXIS USED AN OLDER DEFINITION OF G an L WHICH DID NOT HAVE THE PREFACTOR OF $\gamma_i$ ...}
%\label{fig:wiggles1}
%\end{figure}

\begin{figure}[h!]
\centering
\includegraphics[width = \columnwidth , height = 5 in]{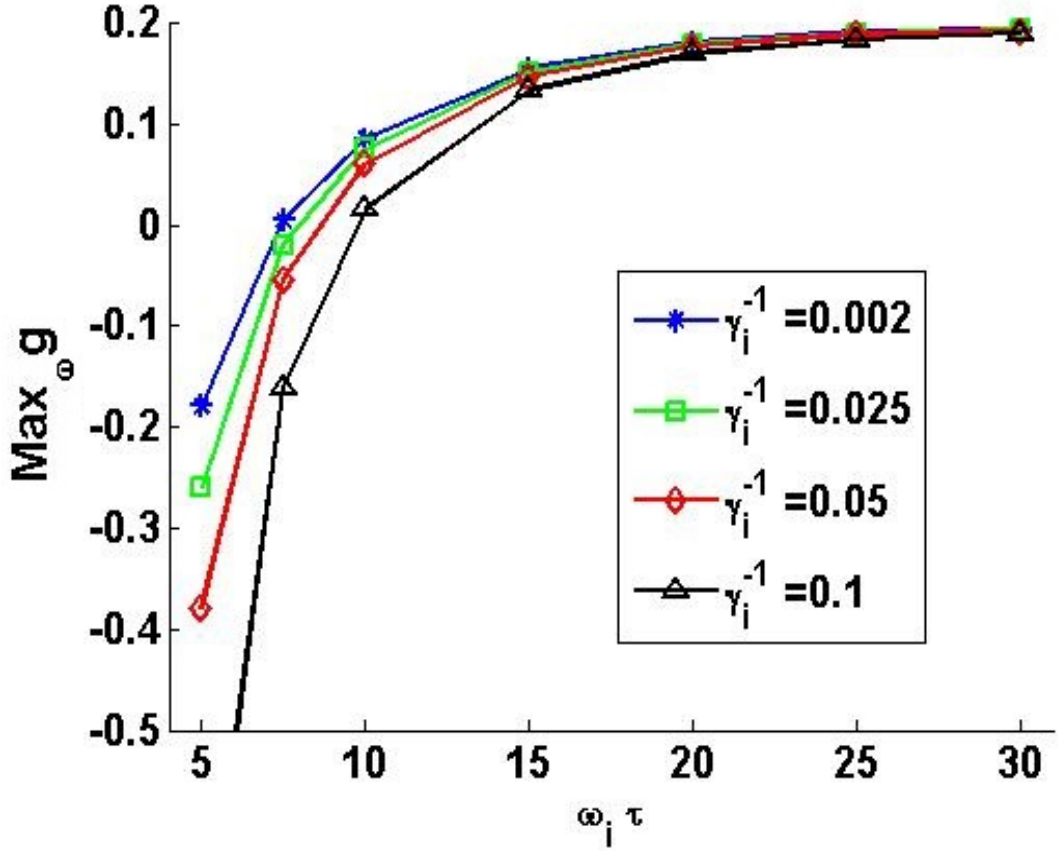}
\caption{Gain maximized over frequency as a function of slowing down time $\omega_i \tau$ and initial normalized half-bandgap $\gamma_i^{-1}$.}
\label{fig:maxes1}
\end{figure}

The gain curves of the type shown in Fig. \ref{fig:wiggles1} and \ref{fig:bandgapvariation1} have a series of local maxima as functions of $\omega / \omega_i$. We record for each pair of parameters $\gamma_i^{-1}$ and $\omega_i \tau$ the maximum value of gain and plot these maxima as functions of $\omega_i \tau$ in Fig. \ref{fig:maxes1}.

\newpage \

\begin{figure}[h!]
\centering
\includegraphics[width = \columnwidth , height = 5 in]{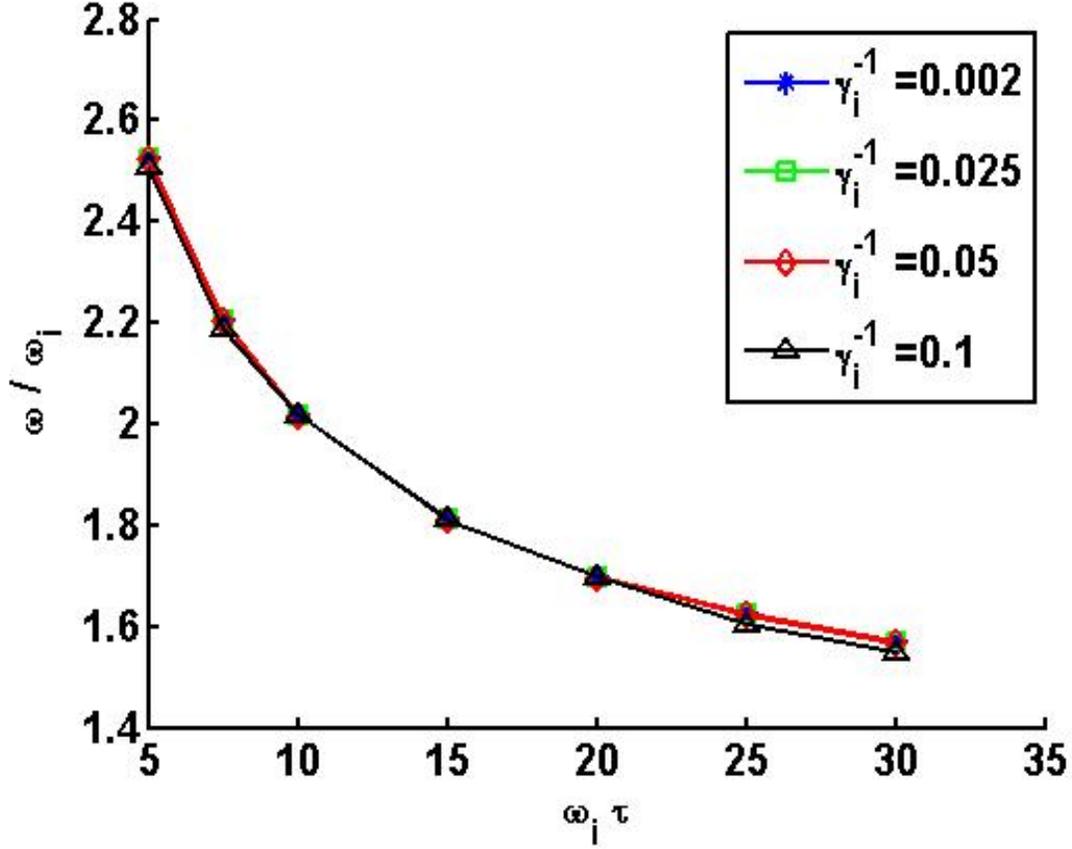}
\caption{Dimensionless frequency at which the gain peaks in Fig. (\ref{fig:bandgapvariation1}) and (\ref{fig:maxes1}) occur }
\label{fig:positionofmaxes1}
\end{figure}

Two things are apparent in Fig. \ref{fig:maxes1}. First, the maximum in gain is relatively insensitive to the half bandgap energy once it is less than $\gamma_i^{-1} \leq 0.1$, and insensitive to slowing down time once it reaches $\omega_i \tau > 20$. We also notice that a slowing down time $\omega_i \tau > 15$ is required for sufficient gain. Figure \ref{fig:positionofmaxes1} shows the frequency corresponding to the maximum gain points of Fig. \ref{fig:maxes1} . We see that for $\omega_i \tau = 20$ the maximum gain occurs for a frequency $\omega / \omega_i \simeq 1.7$. The results of Figs. \ref{fig:maxes1} and \ref{fig:positionofmaxes1} will be used in the next section to determine the optimum dimensional parameters for observing gain.

%
%
%Because the damping model being used here is too simplistic to account for the actual scattering processes of the electrons in graphene, we consider the effect of a spread in damping times. For the value $\tau = $ 428 fs, the peak conduction electron gain (i.e. max value of G - L) occurs at a frequency of approx. 12.6 THz. At this frequency, positive conduction electron gain (i.e. G - L \textgreater 0) occurs for values of $\tau$ deviating by up to a tolerance of over $\pm 20 \%$ from the base value of 428 fs (that is, over the range from approx. 342 fs - 514 fs). This means that a distribution of effective damping times spanning this range still can produce net gain.

\subsubsection*{III. Discussion}

\newpage
\begin{figure}[h!]
\centering
\includegraphics[width = \columnwidth , height = 5 in]{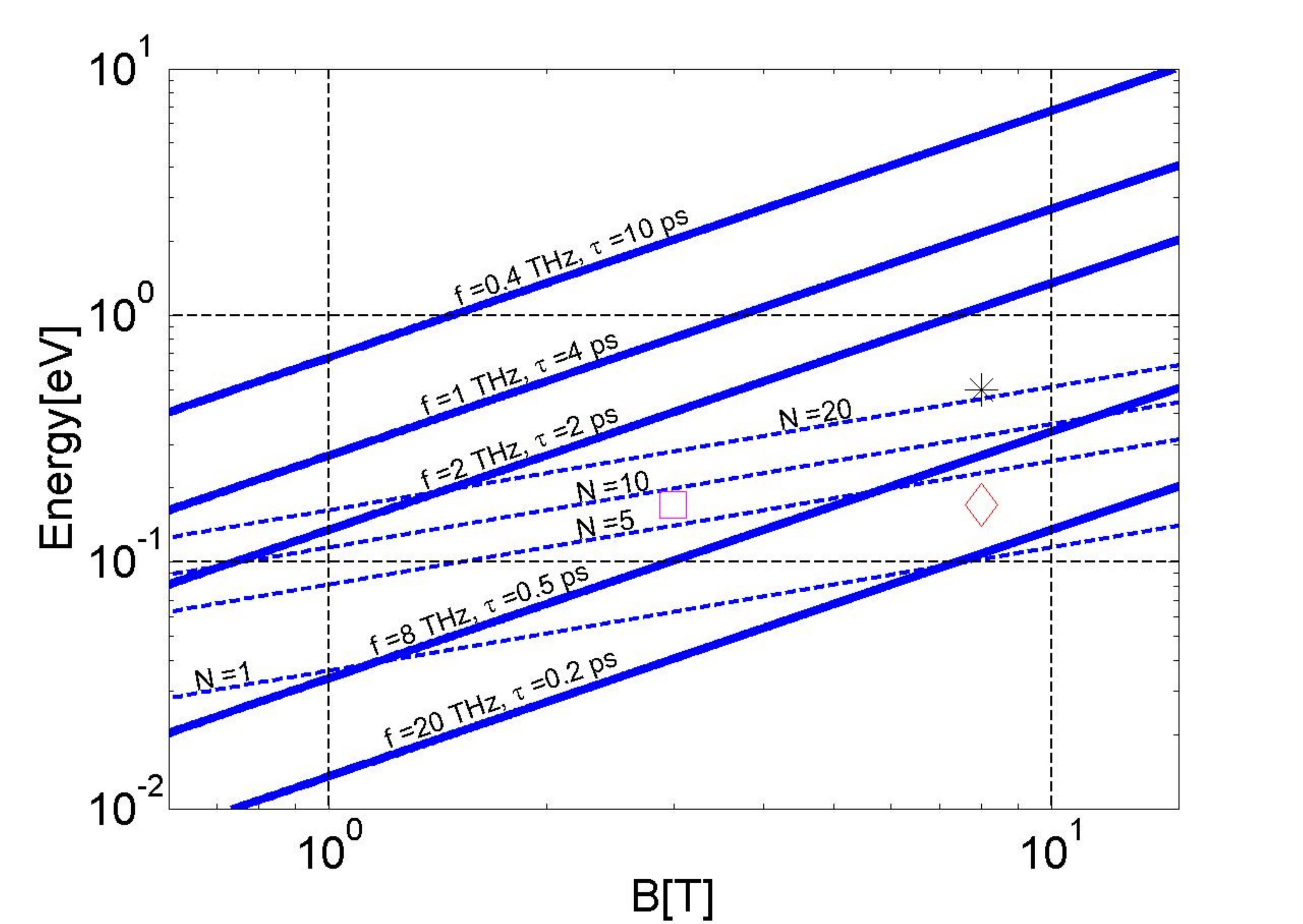}
\caption{Frequency and quantum index as a function of energy and magnetic field (log-log). Magenta square is 3 T magnetic field and 171 meV energy, the other two are 8 T field and 171 meV/500 meV (respectively).}
\label{fig:antonsenengineeringformulaplot1}
\end{figure}

\begin{figure}[h!]
\centering
\includegraphics[width = \columnwidth , height = 5 in]{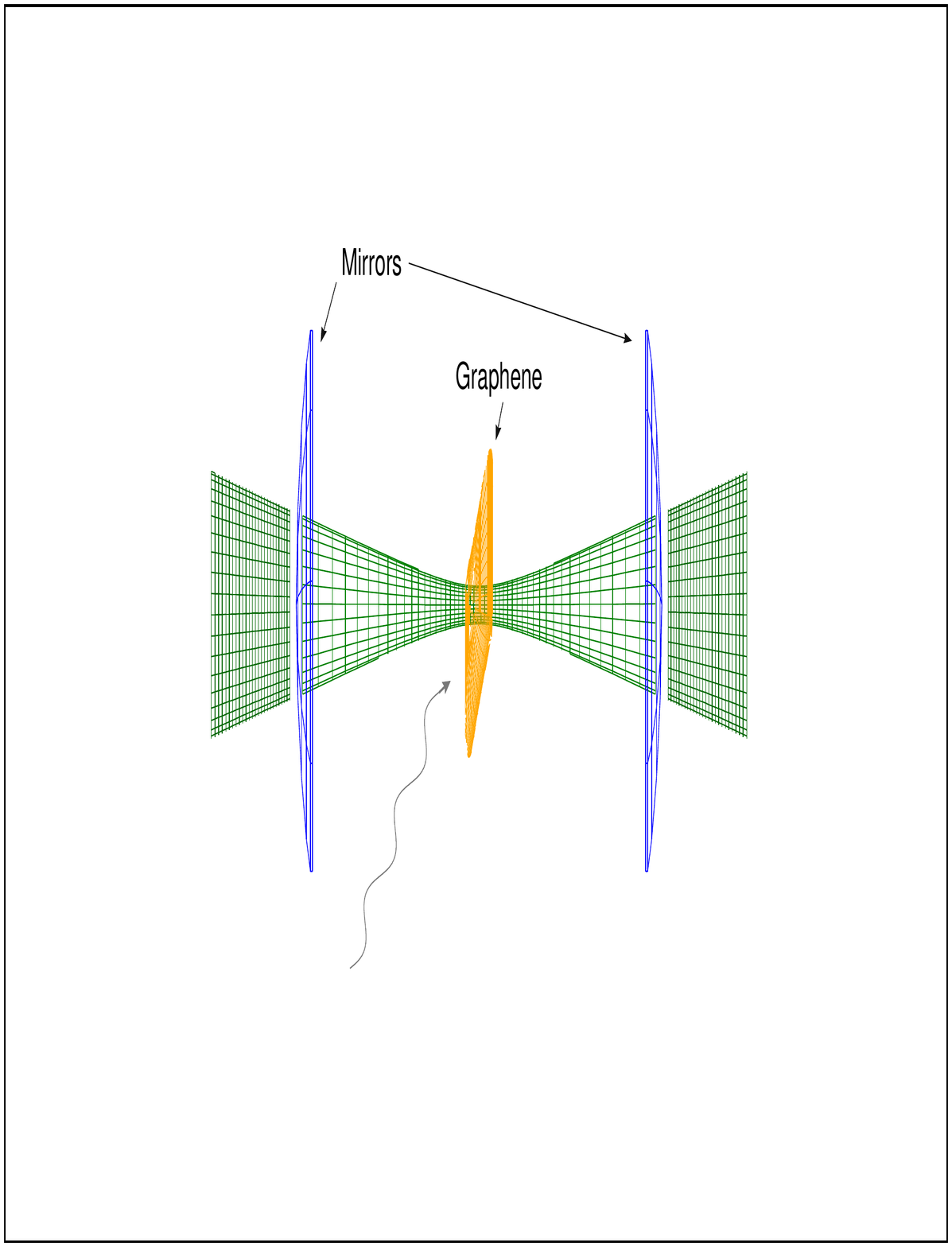}
\caption{Schematic of apparatus using mid-IR laser pumping. Partially transmissive curved mirrors (blue) form a cavity and contain a beam of THz radiation (green). This is amplified by the graphene (yellow-orange) which is pumped by mid-infrared radiation (gray wavy line) from a laser (not shown). The cavity is symmetric and gives two identical output beams (slightly darker green).}
\label{fig:cavity1}
\end{figure}

\indent \indent If THz gain is to be observed in experiments, or if a device producing THz radiation is to be constructed, parameters must be found such that the amplification of the THz field as expressed in Eq. (\ref{eqn:new13}) is sufficient to overcome the intrinsic losses. These are usually in the range of a few percent. In the case of the whole oscillator, the amplification would have to exceed, in addition, the losses and output coupling associated with the oscillator cavity.

The amplification expressed in Eq. (\ref{eqn:new13}) is the product of two factors: The dimensionless gain, $g = Re (G-L)$, and a dimensionless ionization rate $R$ given by (\ref{eqn:new50}). To be sufficient to overcome all losses, we must have $Rg > \ell$ for amplitude losses $\ell$ . Thus, the dimensionless gain $g$ must be considered along with total cavity losses of all types in order to come up with the minimum value of R needed for the cavity to oscillate. These losses include mirror transmission, mirror scattering and absorption, and valence electron absorption in graphene, assumed to be 2.3 \% of power for a monatomic layer. As an example of cavity loss, a mirror transmission $T$ = 0.0125, mirror absorption/scattering $\alpha$ of 0.01, and graphene absorption $\xi$ of $0.023$, combine to form a cavity single pass {\it power} loss of $1-(1-T-\alpha)(1- \xi)$ = 0.045, or 4.5 \%. The {\it amplitude} loss $ \ell $ for this example is $1-\sqrt{(1-T-\alpha)(1- \xi)}$ = 0.0228, or 2.28 \%. Note that the assumed graphene absorption of 2.3 \% is the standard absorption for a single layer of graphene, although that value is considered in much of the relevant literature (e.g. \cite{paper15}) to be accurate only at higher frequencies than the THz range, and thus, may be a poor approximation at said frequencies.

\noindent  Since the normalized gain will be at best $g \sim 0.15$, it follows from Eq. (\ref{eqn:new13}) that an (extremely large) cavity amplitude loss of $15 \%$ implies a rate which must be in the range $R \sim 1$. For the minimal (graphene only) loss, we must have $R > 0.077$. We can rewrite $R$ as follows,

\begin{align}
R = 5.76 \times 10^2 \frac{(\tau[\rm{ps}])^2}{E[\rm{eV}] \tau_I[\rm{ps}]},
\end{align}

\noindent where $\tau_I = n_o/\dot{n}$ is an average ionization time and $n_0$ is the surface density of valence electrons, $n_0 = 3.82 \times 10^{19} m^{-2}$. The requirement $R > 0.077$ then gives

\begin{align}
\frac{1}{\tau_I[\rm{ps}]} > 1.34 \times 10^{-4} \frac{E[\rm{eV}]}{(\tau[\rm{ps}])^2}.
\end{align}

The ionization time also determines the pumping laser power per unit area absorbed, given by

\begin{align}
I = \frac{2 n_0 E_i}{\tau_I} = 1.643 \times 10^5 \left( \frac{E[\rm{eV}]}{\tau[\rm{ps}]} \right)^2 [\rm{W/cm^2}].
\end{align}

We note that if the constraint $\omega_i \tau = 20$ is imposed then,

\begin{align}
I[\rm{W/cm^2}] = 4.11 \times 10^2 (B[\rm{T}])^2 .
\label{eqn:new51}
\end{align}

  Possible operating points are displayed on Fig. (\ref{fig:antonsenengineeringformulaplot1}) in a plot of the magnetic field vs. electron energy plane. Recall that the electron energy will be half the photon energy of the pumping laser. Two sets of lines are displayed in this plot.  The solid lines are lines of constant frequency as given by Eq. (\ref{eqn:new49}).  The dashed lines are lines of constant $N$ as given by Eq. (\ref{eqn:new48}). The present analysis applies only to points well above the N = 1 line. Using the results of Fig.\ref{fig:wiggles1}, we estimate that N $\textgreater$ 10 is required. The set of lines labeled with frequency values show the energy and magnetic field needed to produce gain at the indicated frequency. Here we have taken the operating frequency to be a factor 1.7 times greater than the initial gyration frequency such that it is given by Eq. (\ref{eqn:new49}). Operation at a specified frequency requires that the slowing down time be sufficiently long such that $\omega_i \tau \geq 20$. Thus, curves of constant operating frequency also correspond to curves of required slowing down time. As far as parameter choices are concerned, high electron energies make the mean free time too short, for example, at electron energies in the 0.5-1 eV range (asterisk in Fig. (\ref{fig:antonsenengineeringformulaplot1}) is at 500 meV), the mean free time is only around 100 fs or less due to the hot electrons losing energy to interband transitions (``impact ionization'')\cite{paper16},\cite{paper17}. For this and other reasons, the parameter choices corresponding to Fig. (\ref{fig:wiggles1}) seem more reasonable and are shown by the diamond on Fig. (\ref{fig:antonsenengineeringformulaplot1}). At this low energy (corresponding to an oscillator operation frequency of 12.6 THz if the magnetic field is 8 T), the energy loss rate should be small because 171 meV is below the threshold for optical phonon emission \cite{paper18}. Note also that magnetic fields can increase intraband relaxation times in graphene \cite{paper19}, \cite{paper20}. The absorbed pump laser power required for this example follows from Eq. (\ref{eqn:new51}) and is $2.63 \times 10^4 \rm{W/cm^2}$. If this power is absorbed in an area whose diameter is 20 wavelengths at 12.63 THz, the required absorbed power is $\sim$ 47 W. The square shows a reduced magnetic field that corresponds to a higher quantum level number, which can be considered to make sure the classical formulas are applicable.

%  ,             One example would be B = 8 T, f = 4 THz, E = 500 meV, $\tau$ = 1 ps is indicated by the dot on the 4 THz line. The absorbed pump laser power required for this example follows from Eq. (\ref{eqn:new51}) and is $4.67 \times 10^4 W/cm^2$. If this power is absorbed in an area whose diameter is 20 wavelengths at 12.63 THz, the required absorbed power is ~83 W. This would most likely require operating in pulse mode to keep heating of the graphene at an acceptable level. However, at electron energies in the 1 eV range, the mean free time is only around 100 fs or less due to the hot electrons losing energy to interband transitions (``impact ionization''). \cite{paper7},\cite{paper8}
%
%  By contrast, the dot open circle on the same plot represents the parameter choices corresponding to Fig. (\ref{fig:wiggles1}). At this low energy, the energy loss rate should be small because 171 meV is below the threshold for optical phonon emission \cite{paper9}

Up until now we have only considered the single pass amplification of a THz wave incident on a single layer of graphene. If a self-sustaining oscillator is desired, it would be configured as shown in Fig. (\ref{fig:cavity1}). The graphene would be placed between two mirrors that define a Fabry-Perot resonator, and the THz wave would pass repeatedly through the graphene. The THz signal would grow from noise, if the gain were sufficient to overcome losses, $R g > \ell $ where $\ell$ represents the amplitude loss factor per half trip through the resonator. As mentioned, contributing to $\ell$ are the intrinsic losses in the graphene, losses in the mirrors, and any fractional losses due to output coupling.

Two important issues to be addressed in the classical picture are the determination of operating frequency and determination of the saturation level of THz radiation. If the frequency were known, determination of the saturation level could be made by returning to Eqs. (\ref{eqn:new6}) and (\ref{eqn:new7}) and solving them numerically with a prescribed field amplitude $\hat{E}$. The recorded trajectories $p(t-t_B,\theta_0)$ and $\theta(t-t_B,\theta_0)$ would then be inserted in Eq. (\ref{eqn:new5})  and a nonlinear gain would be computed. This calculation should be repeated for different amplitudes until the amplitude was found for which the nonlinear gain balanced the losses.

The determination of the operating frequency will require a treatment of the competition between the different modes of the Fabry-Perot resonator. From Figs. (\ref{fig:wiggles1}) and (\ref{fig:bandgapvariation1}) we see that the gain vs. frequency has a series of peaks. The fractional width of a single peak is about $\Delta f/f \simeq 0.1$. Thus, if the spacing between mirrors is L = 1 cm, the separation in frequency between adjacent modes is $\pi c / L = 90$ GHz. Taking the operating frequency to be 12.6 THz and the gain bandwidth to be 1.26 THz implies that only fourteen modes could have gain. The competition among modes could then be treated via expansion of the field in modes with slowly evolving amplitudes.

Another issue worthy of deeper study is the effects of collisions on the electron motion. We have modeled the effect as a steady slowing down. The collision process may also involve scattering in pitch angle and energy. A simple estimate of the sensitivity of our results to the inclusion of these effects can be made by examining the dependence of gain on slowing down time in Fig. (\ref{fig:wiggles1}). We note that it requires a change in slowing down time from $\omega_i \tau = 20$ to $\omega_i \tau = 30$ to move the positive gain band of frequencies by an amount equal to its width. Thus, to the extent that the additional collision processes can be modeled as less than a 50\% variation in slowing down time the conditions for gain are robust.

\subsection*{Conclusion}

In conclusion, a graphene gyrotron-like oscillator can be described using a cavity and a linear model. In the linear model, gain occurs at some frequencies if the electrons have a long enough slowing down time and are assumed to not undergo large-angle scattering. Results are promising and net gain in the THz frequency regime for the entire oscillator might be possible considering cavity, graphene absorption, and output coupling, but some questionable assumptions were made, particularly concerning which scattering processes of electrons in graphene can be neglected, and how to treat the others.

Of course, this analysis requires caution regarding the usage of classical physics, which is valid only when the electron kinetic energy is a large multiple of the gyration quantum energy, $N \gg 1$. The scattering time is now realistically in accordance with modern experimental values, as the scattering time is distinct from that of thermal electrons and is enhanced by the presence of a magnetic field. This magnetic field increases the carrier lifetime significantly above that which is otherwise observed and reported in the literature.

A more complete analysis would also look at possible quantum corrections, thermal excitation effects, and factor in the electron-hole cross section both for scattering and creation (per carbon atom) by the IR laser. More complex mirrors could also be considered.

\section*{Acknowledgements}

We are pleased to acknowledge discussions with Ed Ott, Thomas Murphy ,
 Martin Mittendorff, and Michael Fuhrer.  This work was partially
 supported by the Naval Research Laboratory (N00173131G018) and the
 Office of Naval Research (N000140911190).

\appendix
\section*{Appendix: Semi-analytical treatment of integrals}

In this section, we aim to explain the behavior seen in Fig. \ref{fig:wiggles1} by making approximations to Equation (\ref{eqn:new18}) so as to make it possess a closed-form solution.

\noindent Equation (\ref{eqn:new18}) with $\tau_A (t^{\prime})$ expanded,

\begin{align}\tag{A1}
G = - \gamma_i \int_0^{\infty} d\hat{t} e^{-i \Delta \overline{\theta}_0 (\hat{t})} \frac{i p_0}{\tau^2 \gamma_0} \int_0^{\hat{t}} dt^{\prime} \left( \frac{\omega_L}{\gamma_0^2} \frac{d \gamma_0}{d p_0} \right) e^{- t^{\prime} / \tau} \int_0^{t ^ {\prime}} dt^{\prime \prime} e^{t^{\prime \prime} / \tau + i \Delta \overline{\theta}_0(t^{\prime \prime})},
\end{align}

\noindent can be re-expressed with the lower endpoint contribution of the innermost integral made explicit:

\begin{align}\tag{A2}
G = \frac{G_2 - G_1}{\tau^2},
\label{eqn:new35}
\end{align}
\noindent with
\begin{align}\tag{A3}
G_1 = \gamma_i \int_0^{\infty} d\hat{t} e^{-i \Delta \overline{\theta}_0 (\hat{t})} \frac{i p_0}{\gamma_0} \int_0^{\hat{t}} dt^{\prime} \left( \frac{\omega_L}{\gamma_0^2} \frac{d \gamma_0}{d p_0} \right) e^{- t^{\prime} / \tau} \int_{-\infty}^{t ^ {\prime}} dt^{\prime \prime} e^{t^{\prime \prime} / \tau + i \Delta \overline{\theta}_0(t^{\prime \prime})}
\end{align}
\begin{align}\tag{A4}
G_2 = \gamma_i \int_0^{\infty} d\hat{t} e^{-i \Delta \overline{\theta}_0 (\hat{t})} \frac{i p_0}{\gamma_0} \int_0^{\hat{t}} dt^{\prime} \left( \frac{\omega_L}{\gamma_0^2} \frac{d \gamma_0}{d p_0} \right) e^{-t^{\prime} / \tau} \int_{-\infty}^0 dt^{\prime \prime} e^{t^{\prime \prime} / \tau + i \Delta \overline{\theta}_0(t^{\prime \prime})}.
\end{align}

\noindent It may now be noted that the innermost integral in $G_2$ is separable from the rest since the bounds of integration are fixed and thus the inner integral is independent of the dummy variables in the outer integrals and acts as a constant with respect to them:

\begin{align}\tag{A5}
G_2 = \left[ \gamma_i \int_0^{\infty} d\hat{t} e^{-i \Delta \overline{\theta}_0 (\hat{t})} \frac{i p_0}{\gamma_0} \int_0^{\hat{t}} dt^{\prime} \left( \frac{\omega_L}{\gamma_0^2} \frac{d \gamma_0}{d p_0} \right) e^{-t^{\prime} / \tau} \right] \left[ \int_{-\infty}^0 dt^{\prime \prime} e^{t^{\prime \prime} / \tau + i \Delta \overline{\theta}_0(t^{\prime \prime})} \right].
\label{eqn:new27}
\end{align}

\noindent Recall that $e^{-i \Delta \overline{\theta}_0 (\hat{t})}$ is rapidly oscillatory except around $\hat{t} = t_{R}$, so the outermost integral in (\ref{eqn:new27}) gets its main contribution from that time. For low frequencies such that $t_{R} << \tau$, we can thus approximately evaluate that integral by invoking $e^{-t^{\prime} / \tau} \simeq 1$ , $p_0(t^{\prime}) \simeq p_0(t) \simeq p_i$ , and $\gamma_0(t^{\prime}) \simeq \gamma_0(t) \simeq \gamma_i$, so we have (using $d \gamma_0 / dp_0 = p_0/((m^{\prime} c^{\prime})^2 \gamma_0) $):

\begin{align}\tag{A6}
\gamma_i \int_0^{\infty} d\hat{t} e^{-i \Delta \overline{\theta}_0 (\hat{t})} \frac{i p_0}{\gamma_0} \int_0^{\hat{t}} dt^{\prime} \left( \frac{\omega_L}{\gamma_0^2} \frac{d \gamma_0}{d p_0} \right) e^{- t^{\prime} / \tau}
\end{align}
\begin{align}\tag{A7}
\simeq \gamma_i \int_0^{\infty} d\hat{t} e^{-i \Delta \overline{\theta}_0 (\hat{t})} \times \frac{i p_i}{\gamma_i} \int_0^{t_{R}} dt^{\prime} \left[ \frac{\omega_L}{\gamma_i^2} \frac{p_i}{(m^{\prime} c^{\prime})^2 \gamma_i} \right]
\end{align}

\begin{align}\tag{A8}
= \gamma_i \int_0^{\infty} d\hat{t} e^{-i \Delta \overline{\theta}_0 (\hat{t})} \frac{i p_i}{\gamma_i} \left( t_{R} \right) \left[ \frac{\omega_L}{\gamma_i^2} \frac{p_i}{(m^{\prime} c^{\prime})^2 \gamma_i} \right].
\label{eqn:new29}
\end{align}

\noindent From (\ref{eqn:new29}) and (\ref{eqn:new27}),

\begin{align}\tag{A9}
G_2 \simeq -K_0 \int_0^{\infty} d\hat{t} e^{-i \Delta \overline{\theta}_0 (\hat{t})},
\label{eqn:new32}
\end{align}

\noindent where $K_0$ is an overall constant factor.

Now the integral in (\ref{eqn:new32}) may be evaluated by noting that the complex exponential is rapidly oscillatory except around
$\hat{t} = t_{R}$ so that it should not make a significant difference whether the lower limit of the integral is at $\hat{t} = 0$ or at $\hat{t} = -\infty$. Thus,

\begin{align}\tag{A10}
\int_0^{\infty} d\hat{t} e^{-i \Delta \overline{\theta}_0 (\hat{t})} \simeq \int_{-\infty}^{\infty} d\hat{t} e^{-i \Delta \overline{\theta}_0 (\hat{t})}.
\label{eqn:new33}
\end{align}

\noindent Now we note that the quadratic phase approximation (\ref{eqn:new52}) may be substituted into (\ref{eqn:new33}), giving

\begin{align}\tag{A11}
\int_0^{\infty} d\hat{t} e^{-i \Delta \overline{\theta}_0 (\hat{t})} \simeq \int_{-\infty}^{\infty} d\hat{t} e^{-i \phi_{R}} e^{-i \frac{\dot{\Omega}}{2} \left( \hat{t} - t_{R} \right) ^2} \\ \tag{A12}
= e^{-i \phi_{R}} \sqrt{\frac{2}{\dot{\Omega}}} \int_{-\infty}^{\infty} e^{-i u^2} du,
\end{align}

\noindent where $\phi_{R}$ is defined by (\ref{eqn:new41}), and we have made the $u$-substitution $u = \left( \hat{t} - t_{R} \right) \sqrt{\dot{\Omega}/2}$. This may be evaluated using Fresnel integrals, giving

\begin{align}\tag{A13}
\int_0^{\infty} d\hat{t} e^{-i \Delta \overline{\theta}_0 (\hat{t})} \simeq e^{-i \left( \phi_{R} + \frac{\pi}{4} \right)} \sqrt{\frac{2 \pi}{\dot{\Omega}}}.
\label{eqn:new34}
\end{align}

\noindent This result (\ref{eqn:new34}) can be inserted in (\ref{eqn:new32}):

\begin{align}\tag{A14}
G_2 \simeq -K_0 e^{-i \left( \phi_{R} + \frac{\pi}{4} \right)} \sqrt{\frac{2 \pi}{\dot{\Omega}}}.
\label{eqn:new36}
\end{align}

\noindent The gain, $G$ may now be evaluated using (\ref{eqn:new35}) with the approximation $G_1 << G_2$ and the result (\ref{eqn:new36}):

\begin{align}\tag{A15}
G \simeq -K_1 e^{-i \left( \phi_{R} + \frac{\pi}{4} \right)},
\label{eqn:new37}
\end{align}

\noindent with $K_1$ being a new overall constant which absorbs some other terms. Thus, $g \textgreater 0$ is only expected to occur when $cos\left( \phi_{R} + \pi / 4 \right) \textless 0 $, in excellent agreement with the numerically integrated result displayed in Fig. \ref{fig:wiggles1}. Note that this crude approximation works well when at frequencies near to, but slightly above, $\omega_L / \gamma_i$ . The value at lower frequencies is an unphysical artifact since $t_{R}$ does not exist. At frequencies much above $\omega_L / \gamma_i$, the approximation performs poorly due to the fact that the assumptions $p_0(t_{R})  \simeq p_i$ and $\gamma_0(t_{R}) \simeq \gamma_i$ no longer hold.

\newpage

%
%
%\subsubsection*{Parameter values unless otherwise specified}
%
%\
%\
%
%\begin{tabular}{| r @{ = } l |} \hline
%mirror transmission T & 0.0125 \\
%total cavity loss $\zeta$ & 0.045 \\
%Graphene absorption $\xi$ & 0.023 \\
%Numerical integration written in & MATLAB R2011a \\
%\hline
%\end{tabular}

%\begin{figure}[h!]
%\centering
%\includegraphics[width=.75\textwidth]{newgain1.eps}
%\caption{Gain with short damping time using linear model. NOTE THAT THE Y AXIS USED AN OLDER DEFINITION OF G an L WHICH DID NOT HAVE THE PREFACTOR OF $\gamma_i$ ...}
%\end{figure}
%
%\newpage\
%Here, we see that there are several frequency bands with $G-L > 0$ with $\tau \simeq 943 fs$.

\newpage\
%\begin{figure}[h!]
%\centering
%\includegraphics[width=.975\textwidth]{newfig9b.eps}
%\caption{Gain with various short damping times using linear model, versus frequency, both in dimensionless form. The physical damping times $\tau$ are chosen to be exactly 300 fs, 600 fs, and 900 fs for the blue, green, and red curves, respectively. Half-band-gap $m^{\prime} c^{\prime 2}$ is approximately 0.1 meV, initial energy is 42.77 meV, and $p_i/m^{\prime}c^{\prime}$ = 425 which is approximately the same as the maximum value shown in the other plots. Upper x-axis shows physical frequency. The solid portions of each curve indicate where $cos(\phi_R + \pi / 4) < 0$, where $\phi_R$ is defined in Eqs. (\ref{eqn:new41}) and (\ref{eqn:new44}).}
%\end{figure}

%\newpage\

% To be more explicit about losses, the mirror parameters’ relation to each other, and cavity performance is also
% included via the following equations.
% $CL=1-[(1-MT-MA)(1-GA)]$
% where $CL$ is the cavity loss,
% $MT$ is the mirror Transmission coefficient,
% $MA$ is the mirror Absorption coefficient,
% and $GA$ is the graphene Absorption coefficient.

% and we take the new fields E parallel and E perp to be slowly varying in time compared to the gyration at
% $\omega_{Larmor}$

\end{document}